\documentclass{elsart}

 \usepackage{graphicx}

\usepackage{amssymb}

\newcommand{\lsim}{\,{\buildrel < \over {_\sim}}\,}
\newcommand{\gsim}{\,{\buildrel > \over {_\sim}}\,}
\newcommand{\sqrtsNN}{\sqrt{s_{\scriptscriptstyle{{\rm NN}}}}}
\newcommand{\av}[1]{\left\langle #1 \right\rangle}

\newcommand{\gev}{\mathrm{GeV}}

\newcommand{\cm}{\mathrm{cm}}

\newcommand{\mb}{\mathrm{mb}}
\newcommand{\PbPb}{\mbox{Pb--Pb}}

\newcommand{\RAA}{R_{\rm AA}}
\newcommand{\RCP}{R_{\rm CP}}
\newcommand{\pt}{p_T}

\renewcommand{\d}{{\rm d}}

\newcommand{\Kzs}{{\rm K^0_S}}
\newcommand{\La}{\Lambda}
\newcommand{\Al}{\overline\Lambda}

\begin{document}

\begin{frontmatter}



\title{Central-to-peripheral nuclear modification factors
in Pb--Pb collisions at $\sqrtsNN=$ 17.3~GeV }

{\large NA57 Collaboration}

\author[k]{F.~Antinori},
\author[e]{P.A.~Bacon},
\author[f]{A.~Badal\`a},
\author[f]{R.~Barbera},
\author[a]{A.~Belogianni},
\author[e]{I.J.~Bloodworth},
\author[h]{M.~Bombara},
\author[b]{G.E.~Bruno\thanksref{corresp}},
\author[e]{S.A.~Bull},
\author[b]{R.~Caliandro},
\author[g]{M.~Campbell},
\author[g]{W.~Carena},
\author[g]{N.~Carrer},
\author[e]{R.F.~Clarke},
\author[k]{A.~Dainese\thanksref{corresp}},
\author[b]{D.~Di~Bari},
\author[n]{S.~Di~Liberto},
\author[g]{R.~Divi\`a},
\author[b]{D.~Elia},
\author[e]{D.~Evans},
\author[p]{G.A.~Feofilov},
\author[b]{R.A.~Fini},
\author[a]{P.~Ganoti},
\author[b]{B.~Ghidini},
\author[o]{G.~Grella},
\author[d]{H.~Helstrup},
\author[d]{K.F.~Hetland},
\author[j]{A.K.~Holme},
\author[f]{A.~Jacholkowski},
\author[e]{G.T.~Jones},
\author[e]{P.~Jovanovic},
\author[e]{A.~Jusko},
\author[r]{R.~Kamermans},
\author[e]{J.B.~Kinson},
\author[g]{K.~Knudson},
\author[p]{V.~Kondratiev},
\author[h]{I.~Kr\'alik},
\author[i]{A.~Krav\v{c}\'akov\'a},
\author[r]{P.~Kuijer},
\author[b]{V.~Lenti},
\author[e]{R.~Lietava},
\author[j]{G.~L\o vh\o iden},
\author[b]{V.~Manzari},
\author[n]{M.A.~Mazzoni},
\author[n]{F.~Meddi},
\author[q]{A.~Michalon},
\author[k]{M.~Morando},
\author[e]{P.I.~Norman},
\author[f]{A.~Palmeri},
\author[f]{G.S.~Pappalardo},
\author[h]{B.~Pastir\v c\'ak},
\author[e]{R.J.~Platt},
\author[k]{E.~Quercigh},
\author[f]{F.~Riggi},
\author[c]{D.~R\"ohrich},
\author[o]{G.~Romano},
\author[g]{K.~\v{S}afa\v{r}\'{\i}k},
\author[h]{L.~\v S\'andor},
\author[r]{E.~Schillings},
\author[k]{G.~Segato},
\author[l]{M.~Sen\'e},
\author[l]{R.~Sen\'e},
\author[g]{W.~Snoeys},
\author[k]{F.~Soramel\thanksref{permaddr}},
\author[a]{M.~Spyropoulou-Stassinaki},
\author[m]{P.~Staroba},
\author[k]{R.~Turrisi},
\author[j]{T.S.~Tveter},
\author[i]{J.~Urb\'an},
\author[r]{P.~van~de~Ven},
\author[g]{P.~Vande~Vyvre},
\author[g]{A.~Vascotto},
\author[j]{T.~Vik},
\author[e]{O.~Villalobos Baillie},
\author[p]{L.~Vinogradov},
\author[o]{T.~Virgili},
\author[e]{M.F.~Votruba},
\author[i]{J.~Vrl\'akov\'a}, 
and~\author[m]{P.~Z\'{a}vada}

\thanks[corresp]{Corresponding authors: 
                 G.E.~Bruno, via Orabona 4, I-70126 Bari (Italy), 
                 {\tt giuseppe.bruno@ba.infn.it};
                 A.~Dainese, via Marzolo 8, I-35131 Padova (Italy), 
                 {\tt andrea.dainese@pd.infn.it}}
\thanks[permaddr]{Permanent address: {\it University of Udine, Udine, Italy}}

\address[a]{\scriptsize Physics Department, University of Athens, 
            Athens, Greece}
\address[b]{\scriptsize Dipartimento IA di Fisica dell'Universit{\`a}
            e del Politecnico and INFN, Bari, Italy} 
\address[c]{\scriptsize Fysisk Institutt, Universitetet i Bergen, Bergen, 
            Norway} 
\address[d]{\scriptsize H{\o}gskolen i Bergen, Bergen, Norway} 
\address[e]{\scriptsize School of Physics and Astronomy, 
            University of Birmingham, Birmingham, UK} 
\address[f]{\scriptsize University of Catania and INFN, Catania, Italy} 
\address[g]{\scriptsize CERN, European Laboratory for Particle Physics, Geneva,
            Switzerland} 
\address[h]{\scriptsize Institute of Experimental Physics, 
            Slovak Academy of Science, Ko\v{s}ice, Slovakia} 
\address[i]{\scriptsize P.J. \v{S}af\'{a}rik University, Ko\v{s}ice, Slovakia} 
\address[j]{\scriptsize Fysisk Institutt, Universitetet i Oslo, Oslo, Norway} 
\address[k]{\scriptsize University of Padua and INFN, Padua, Italy} 
\address[l]{\scriptsize Coll\`ege de France, Paris, France} 
\address[m]{\scriptsize Institute of Physics, Academy of Sciences of the Czech 
            Republic, Prague, Czech Republic} 
\address[n]{\scriptsize University ``La Sapienza'' and INFN, Rome, Italy} 
\address[o]{\scriptsize Dip. di Scienze Fisiche ``E.R. Caianiello''
            dell'Universit{\`a} and INFN, Salerno, Italy} 
\address[p]{\scriptsize State University of St. Petersburg, St. Petersburg, 
            Russia} 
\address[q]{\scriptsize Institut de Recherches Subatomique, IN2P3/ULP, 
            Strasbourg, France} 
\address[r]{\scriptsize Utrecht University and NIKHEF, Utrecht, 
            The Netherlands}

\begin{abstract}
We present central-to-peripheral nuclear modification factors, $\RCP$, for
the $\pt$ distributions of $\Kzs$, $\La$, $\Al$, and negatively charged
particles, measured at central rapidity 
in \mbox{Pb--Pb} collisions at top SPS energy.
The data cover the 55\% most central fraction of the inelastic 
cross section. The $\Kzs$
and $\La$ $\RCP(\pt)$ are similar in shape to those measured at
$\sqrtsNN=200~\gev$ at RHIC, though they are larger in absolute value. 
We have compared our $\Kzs$ $\RCP$ data to a theoretical calculation.
The prediction overestimates the data at $\pt\approx 3$--$4~\gev/c$, 
unless sizeable parton energy loss is included in the calculation.
\end{abstract}


\end{frontmatter}

\setcounter{footnote}{0}


\section{Introduction}
\label{intro}

The quenching of high transverse momentum ($\pt$) particles
in central heavy-ion collisions is one of the main discoveries at BNL-RHIC.
The effect is quantified using the nuclear modification
factor:
\begin{equation}
\RAA(\pt) = {1 \over \av{N_{\rm coll}}_{\rm C}}\times
\frac{\d^2 N_{\rm AA}^{\rm C}/\d\pt\d y}{\d^2 N_{\rm pp}/\d\pt\d y}\,,
\label{eq:raa}
\end{equation}
where $\av{N_{\rm coll}}_{\rm C}$ is the average number of
nucleon--nucleon (NN) collisions for nucleus--nucleus (AA)
collisions in a given centrality class C. The nuclear modification
factor would be equal to unity if the AA collision were a mere
superposition of $N_{\rm coll}$ independent nucleon--nucleon
collisions. In central \mbox{Au--Au} collisions at a c.m.s. energy
per nucleon--nucleon pair of $\sqrtsNN=200~\gev$, the PHENIX and
STAR experiments have measured a suppression by a factor 4--5 with 
respect to unity in
$\RAA$ for $\pt\gsim 5~\gev/c$, independent of the particle
species~\cite{RAA200}. 
A similar suppression at high $\pt$ is observed also in 
the central-to-peripheral nuclear modification factor
\begin{equation}
\RCP(\pt) = {\av{N_{\rm coll}}_{\rm P} \over \av{N_{\rm coll}}_{\rm C}}\times
\frac{\d^2 N_{\rm AA}^{\rm C}/\d\pt\d y}{\d^2 N_{\rm AA}^{\rm P}/\d\pt\d y}\,,
\label{eq:rcp}
\end{equation}
where a class P of peripheral nucleus--nucleus collisions replaces 
the pp reference (see e.g.~\cite{starklaRcp}). 
The measured suppression is interpreted as being due
to energy loss of the hard partons traversing the
high-density QCD medium expected to be formed in high-energy
heavy-ion collisions~\cite{white}. 
Parton energy loss would predominantly occur
via the mechanism of medium-induced gluon
radiation~\cite{gyulassywang,bdmps}.

Preliminary results~\cite{RAA62} from the RHIC run at
$\sqrtsNN=62.4~\gev$ show a $\RAA$ suppression by about a factor 3,
not much smaller than that observed at $\sqrtsNN=200~\gev$. This motivates
the search for parton energy loss effects at even lower c.m.s. energy, i.e.
in \mbox{Pb--Pb} collisions with $\sqrtsNN=17.3~\gev$ at CERN-SPS.
Possible indications of such effects were found by the WA98
Collaboration~\cite{wa98pi0Rcp}, who observed a suppression of the
 $\pi^0$ $\RCP$.
In addition, a recent $\pi^0$ $\RAA$ analysis~\cite{denterria}, using the
WA98 Pb--Pb data~\cite{wa98pi0Rcp} and a parameterization of a wide
compilation of pp
data at similar energy available in the literature, favours a scenario of
significant energy loss.

We investigate the presence
of energy loss effects in \mbox{Pb--Pb} collisions at
$\sqrtsNN=17.3~\gev$ by measuring $\RCP(\pt)$ for $\Kzs$, $\La$
and $\Al$ particles, and for unidentified negatively charged
particles, $h^-$. 
The relative behaviour of the $\RCP(\pt)$
patterns for $\Kzs$ and $\La$ particles is also expected to be sensitive to
parton coalescence effects~\cite{recoth} in the hadronization
dynamics. It has been suggested that these effects occur at RHIC energy (see
e.g.~\cite{sorensen}).

In section~\ref{analysis} the experimental setup, 
the data sets and the analysis procedures are 
described. The $\RCP$ results are presented in section~\ref{results} and
they are compared  in section~\ref{discussion} 
to other experimental results at SPS and RHIC energies as well as
with theoretical calculations.

\section{Apparatus, data sets and analysis}
\label{analysis}

The NA57 apparatus, described in detail in~\cite{na57setup}, was
designed to study the production of strange and multi-strange
particles in fixed-target heavy-ion collisions by reconstructing
their weak decays into final states containing charged particles only.
Tracks are reconstructed in the 30 cm-long silicon
telescope: an array of pixel detector planes with a cross
section of $5\times 5~\cm^2$ placed inside a 1.4~T magnetic field
normal to the beam direction. The telescope is inclined in the
non-bending plane by a 40~mrad angle relative to the beam
line and it points to the target located 60~cm from
the first detector plane.
The acceptance covers about half a unit in
rapidity at central rapidity and transverse momentum larger than
about $0.5~\gev/c$. For the \PbPb~runs, the centrality trigger,
based on charged multiplicity,
was set so as to select approximately the most central 60\% of the inelastic
collisions. 

The results presented
in this letter are based on the analysis of 
data collected during the 1998 (about
$2.1\cdot 10^8$ events) and 2000 (about $2.5\cdot 10^8$ events) 
\PbPb~runs with beam momentum of 158~$A~\gev/c$.

\begin{table}[!t]
\caption{Average number of participants and of NN collisions
         with their systematic errors, as a function of 
         centrality.}
\label{tab:centrality}
\begin{center}
\begin{tabular}{ccc}
\hline
Class (\% $\sigma^{\rm Pb-Pb}_{\rm inel}$) & $\av{N_{\rm part}}$ & $\av{N_{\rm coll}}$ \\
\hline
0--5.0\% & $345.3 \pm 1.7$ & $779.2 \pm 26.6$ \\
10.0--20.0\% & $214.7 \pm 5.8$ &  $421.7 \pm 26.1$ \\
20.0--30.0\% & $143.0 \pm 6.6$ & $247.7 \pm 21.5$ \\
30.0--40.0\% & $92.6 \pm 6.4$ &  $140.5 \pm 16.2$ \\
40.0--55.0\% & $49.5 \pm 5.0$  & $63.8 \pm 9.8$ \\
\hline
\end{tabular}
\end{center}
\end{table}

The collision centrality is determined using the charged particle
multiplicity $N_{\rm ch}$ in the pseudorapidity range $2<\eta<4$,
sampled by the microstrip silicon detectors (MSD) 
as described in~\cite{wounded,mult2004}.
$N_{\rm ch}$ is related to the centrality assuming
$N_{\rm ch}=q\cdot N_{\rm part}^\alpha$ 
(a modified Wounded Nucleon model)~\cite{wounded}, where
$N_{\rm part}$ is the number of participants, i.e. nucleons participating
in the primary nucleon--nucleon collisions, 
estimated from the Glauber model~\cite{glauber}. 
The $N_{\rm ch}$ distribution is
well described by a Wounded Nucleon model fit with
$\alpha=1.0$~\cite{mult2004}.
The inelastic cross section extracted from the fit~\cite{wounded},
using an inelastic non-diffractive NN cross section 
$\sigma^{\rm NN}_{\rm inel}=30~\mb$~\cite{carroll,gottgens},
is $\sigma^{\rm Pb-Pb}_{\rm inel}=7.26$~b. Five centrality classes
are defined, with $N_{\rm ch}$ limits corresponding to given
fractions of  $\sigma^{\rm Pb-Pb}_{\rm inel}$. For each class the average
values of $N_{\rm part}$ and $N_{\rm coll}$ are calculated from the
Glauber model, with Pb nucleus Woods-Saxon density-profile parameters as
given in~\cite{ws}. In Table~\ref{tab:centrality}
we present the definitions of the five centrality classes along
with the corresponding values of $\av{N_{\rm part}}$ and $\av{N_{\rm coll}}$
with their systematic errors, estimated by varying the
$\alpha$ parameter in the range [1.0,1.1], the Woods-Saxon parameters
within their tabulated uncertainties and the inelastic 
non-diffractive NN cross section at $\sqrt{s}\simeq 17~\gev$ 
within its systematic error estimated to
be 1.5\%~\cite{carroll,gottgens}\footnote{The $\av{N_{\rm part}}$ values 
given in Table~\ref{tab:centrality} for the most central and the most 
peripheral classes differ slightly from those published in~\cite{mult2004}
due to a refined treatment of the trigger effect in the fit of
the multiplicity distribution done for the present analysis.}. Note, however, 
that the contribution due to the uncertainty on $\sigma^{\rm NN}_{\rm inel}$
is negligible for $\av{N_{\rm part}}$, while for $\av{N_{\rm coll}}$
it ranges from 1.5\% in the most central class to 1.1\% 
in the most peripheral, but it cancels out in the ratio 
$\av{N_{\rm coll}}_{\rm P}/\av{N_{\rm coll}}_{\rm C}$ used for $\RCP$.

\begin{figure}[!t]
\begin{center}
   \includegraphics[width=\textwidth]{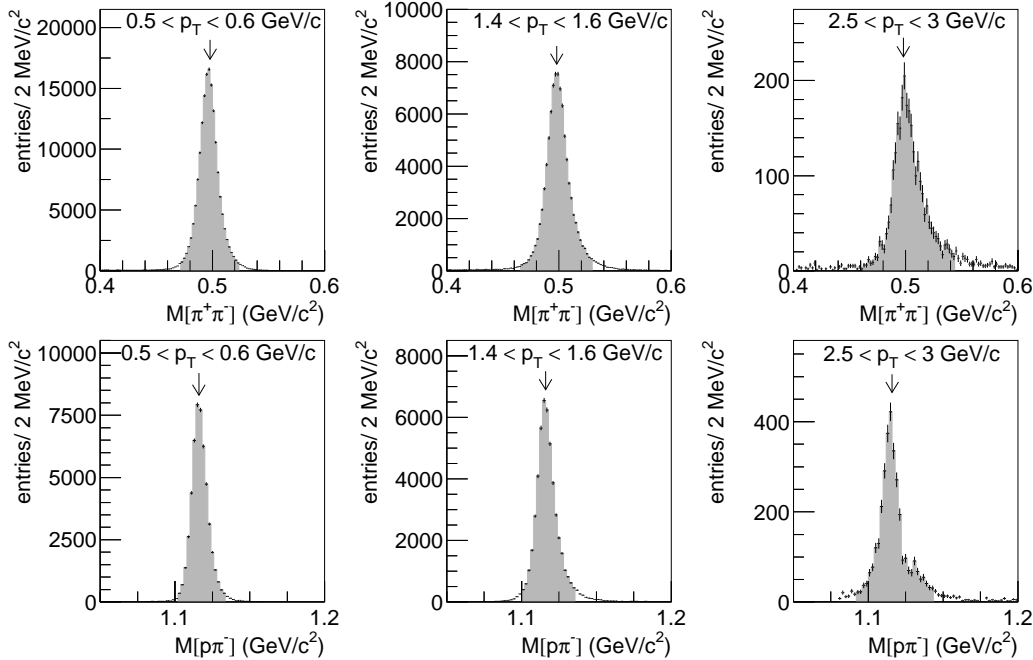}
   \caption{Invariant mass distributions for $\Kzs$ (upper row) and
            $\Lambda$ (lower row) candidates in different $\pt$ ranges.
            Arrows indicate the nominal masses of the two particles and 
            shaded areas indicate the ranges considered for the analysis.
            No centrality selection is applied.}
   \label{fig:masses}
\end{center}
\end{figure}

Strange particles are reconstructed using their decay channels
into charged particles:
$\Kzs \to \pi^+\pi^-$, $\La\to\pi^-{\rm p}$, and
$\Al\to\pi^+\overline{\rm p}$.
The selection procedure is described in detail
in~\cite{wa97v0sele,na57v0sele}. The main criteria are the following:
(a) the two decay tracks are compatible with the hypothesis of having a common
    origin point;
(b) the reconstructed secondary vertex is well separated in space from the
      target;
(c) the reconstructed candidate points back to the primary vertex position.
For $\Kzs$, the two decay tracks are required to miss the
interaction vertex by applying a $\pt$-dependent cut on
the product of their impact parameters
(distances of closest approach to the interaction vertex in the
bending plane). 
Ambiguities among $\Kzs$, $\La$ and $\Al$ are eliminated 
by means of cuts~\cite{wa97v0sele} on the
Podolanski-Armenteros plot~\cite{armenteros}.
The final invariant mass
distributions for $\Kzs$ and $\La$ candidates in different
$\pt$ ranges are shown in Fig.~\ref{fig:masses}.
 The shaded areas correspond to the
windows used for this analysis;
the size of these windows is smaller at low $\pt$, where the
 invariant mass resolution is better.
The statistics amounts to
$1.8\cdot 10^6$ $\Kzs$, $0.7\cdot 10^6$ $\La$, and $0.1\cdot 10^6$ $\Al$.
We have estimated and subtracted the combinatorial background
using the event-mixing technique on a subsample of events representative of 
the full statistics~\cite{na57v0sele}.
The background fraction in the selected mass window is negligible for our 
most peripheral class (40--55\%) and it increases going to more central 
classes, where it also increases from low to high transverse momentum. 
In the worst case, in class 0--5\%, the subtracted background 
fraction amounts to:
$(6\pm 2)\%$ for $\Kzs$ with $\pt>2.5~\gev/c$; $(10\pm 3)\%$ for $\La$
with $\pt>2.5~\gev/c$; $(8\pm 3)\%$ for $\Al$ with $\pt>2~\gev/c$.

Negatively charged particles, $h^-$, are selected from a sample of good-quality
tracks (clusters in more than 80\% of the telescope planes,
less than 30\% of the clusters shared with other tracks) using 
an impact parameter cut to ensure they
come from the interaction vertex. The
residual contamination of secondary tracks (decay products of 
weakly-decaying particles) has been estimated to be of about 2\%,
independent of $\pt$,
for the most central class of events, and lower
for the other classes.
The statistics amounts to $10^8$ particles.

\begin{figure}[!t]
\begin{center}
   \includegraphics[width=0.5
   \textwidth]{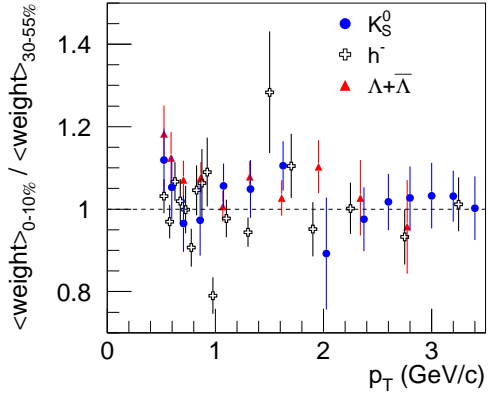}
   \caption{(colour online) 
            Ratio of the average correction weights for a sample
            of central events (0--10\%) and a sample of peripheral events
            (30--55\%), as a function of $\pt$ and of the particle species.}
   \label{fig:weights}
\end{center}
\end{figure}

We calculate $\RCP(\pt)$, Eq.~(\ref{eq:rcp}), using $\pt$
distributions which are unweighted, 
i.e.\,not corrected for geometrical acceptance 
and reconstruction/selection efficiency.
For other analyses, in particular for 
the measurement of the $\pt$-integrated strange particle production 
yields~\cite{na57yields},
we adopted a procedure, specifically developed for rarer signals like $\Xi$
and $\Omega$,
in which every selected particle was assigned a correction weight,
calculated on the basis of a Monte Carlo simulation~\cite{na57v0sele}.
This correction is very time consuming and,
for the more abundant signals ($h^-$, $\Kzs$ and $\La$), it 
was calculated only for a representative subsample of the available statistics.
For the present study, we have verified that these weights
do not depend on the event
centrality over the full transverse momentum range covered 
(see Fig.~\ref{fig:weights}).
Using unweighted spectra results in a negligible
systematic error on $\RCP(\pt)$ 
as compared to the other contributions discussed in the following.

\section{Results}
\label{results}

We use class 40--55\% as the reference peripheral class in the
denominator of $\RCP$, see Eq.~(\ref{eq:rcp}), and vary the
`central' class in the numerator from 0--5\% to 30--40\%.
We estimated possible residual systematic 
effects by comparing the $\RCP$ results
corresponding to the year 1998 sample and to the year 2000 sample,
which were also processed with different versions of the reconstruction
software. We find compatible results, with 
point-to-point differences smaller than 10\% for
$h^-$, $\Kzs$ and $\La+\Al$. We assign `reconstruction' 
systematic errors of 5\% for all four species.
For strange particles, a small contribution to the systematic errors has
been recognized as due to the procedure used to remove fake tracks, 
which yield duplicated
candidates, particularly in central collisions that have higher track 
density\footnote{For the $h^-$ analysis, 
fake tracks were avoided by imposing tighter track-quality conditions.}.
For $\Kzs$, $\La$ and $\Al$,
the resulting error, estimated to be at most 3\% at high $\pt$ for the ratio
0--5\%/40--55\% and smaller for the other ratios, has been added 
in quadrature to the `reconstruction' systematic error and to the 
systematic error due to background subtraction, 
which is also at most 3\%.

\begin{figure}[!t]
\begin{center}
   \includegraphics[width=\textwidth]{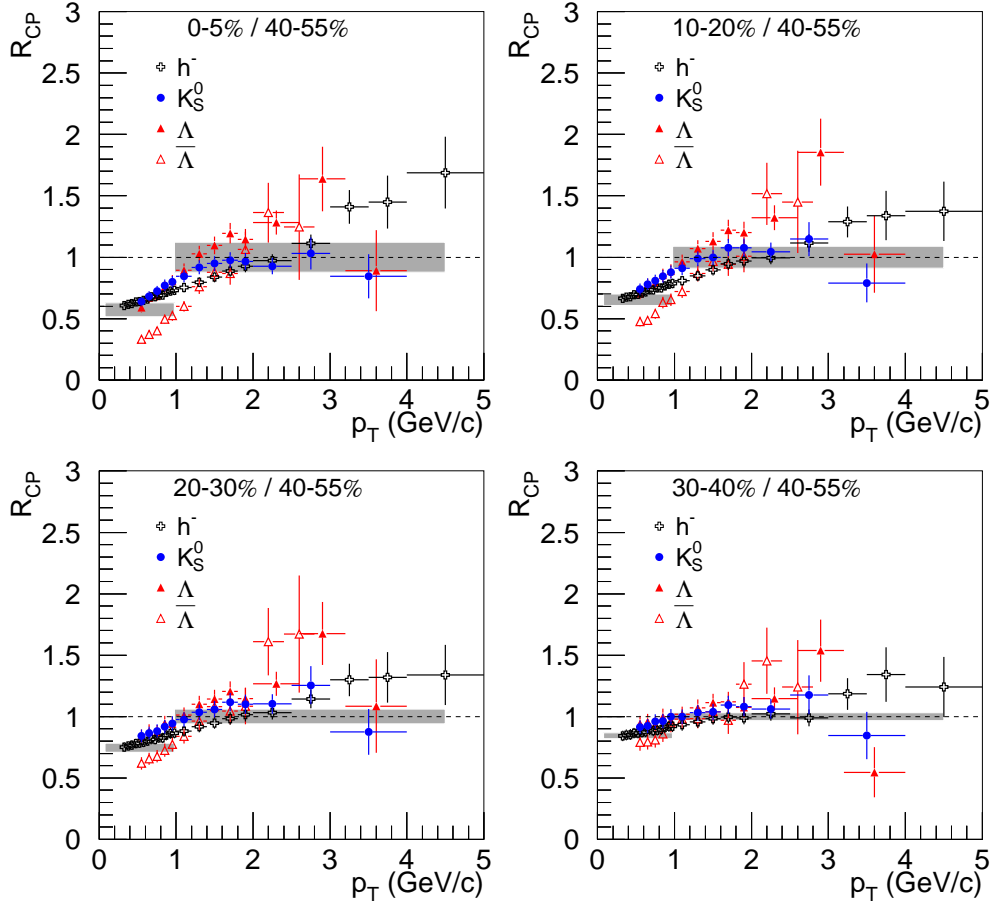}
   \caption{(colour online) Centrality dependence of $\RCP(\pt)$
            for $h^-$, $\Kzs$, $\La$ and $\Al$
            in \PbPb~collisions at $\sqrtsNN=17.3~\gev$. Shaded bands centered
            at $\RCP=1$ represent the systematic error due to the uncertainty 
            in the ratio of the
            values of $\av{N_{\rm coll}}$ in each class; shaded
            bands at low $\pt$ represent the values expected for
            scaling with the number of participants, together with their 
            systematic error.}
   \label{fig:Rcp}
\end{center}
\end{figure}

Figure~\ref{fig:Rcp} shows our $\RCP(\pt)$ results for four
different `central' classes (C). The error bars are obtained as a
quadratic sum of the statistical errors and the
$\pt$-dependent systematic errors ($< 7\%$). The
shaded bands centered at $\RCP=1$ represent the $\pt$-independent systematic
error due to the uncertainty in the ratio 
$\av{N_{\rm coll}}_{\rm P}/\av{N_{\rm coll}}_{\rm C}$, 
while the shaded bands at low $\pt$
represent the $\RCP$ values corresponding to
$N_{\rm part}$-scaling, with the band indicating the
systematic error due to the uncertainty in the ratio 
$\av{N_{\rm part}}_{\rm C}/\av{N_{\rm part}}_{\rm P}$. 
We first focus on the 0--5\%/40--55\% ratio. At
low $\pt$ ($\approx 0.5~\gev/c$), the $h^-$, $\Kzs$ and $\La$ patterns are
compatible with $N_{\rm part}$-scaling, while the $\Al$ points 
are clearly below.
As $\pt$ increases, $\RCP$ for the $\Kzs$ approaches one; 
$\RCP$ for the $\La$ behaves differently from the $\Kzs$ above 
$\pt\approx 1~\gev/c$,
reaching a value of about 1.5, as does the $\Al$. For $\pt$ values below
$2~\gev/c$, the $\RCP$ for negative particles is dominated by negative pions,
and stays below the corresponding values for $\Kzs$.
At higher $\pt$ the $h^-$ $\RCP$ lies between those for $\Kzs$ and $\La$;
in this region 
there may be significant $\rm K^-$ and $\overline{\rm p}$ contributions
in the sample of negative particles (note that
${\rm \overline p}/\pi^- \simeq 0.8$ for 
$\pt\gsim 3~\gev/c$ in central \mbox{Au--Au}
collisions at $\sqrtsNN=200~\gev$~\cite{ptopiPHENIX}).
For less central collisions,
the $\Kzs$ $\RCP$ exhibit a small
enhancement in the range $\pt\gsim 1.2~\gev/c$, while
within errors the $\RCP$ for the other particles do not vary with respect to
0--5\%/40--55\% $\RCP$. 

The difference between the $\La$ and the $\Al$ at low $\pt$, already 
observed in the centrality  
dependence of the $\pt$-integrated yields in \mbox{Pb--Pb} 
collisions~\cite{na57qm04}, is pronounced.  
On the basis of the symmetry of the apparatus and of the  
signal extraction procedure,  
we are confident that the effect is not caused by an experimental bias.  
It may be due to a centrality-dependent absorption effect of $\Al$ 
in a nucleon-rich environment. In this respect, note 
that at SPS energy the $\Al/\La$ ratio is measured to be 
lower by a factor about 2
in \mbox{lead--lead} with respect to sulfur-induced 
collisions~\cite{wa94}
and the $\Al$ yield per participant is measured to be lower 
by a factor $0.71 \pm 0.05$
in \mbox{p--Pb}~\cite{wa97v0sele} with respect to \mbox{p--Be} 
collisions~\cite{na57qm04}.

\section{Comparisons and discussion}
\label{discussion}

\begin{figure}[!b]
\begin{center}
   \includegraphics[width=.9\textwidth]{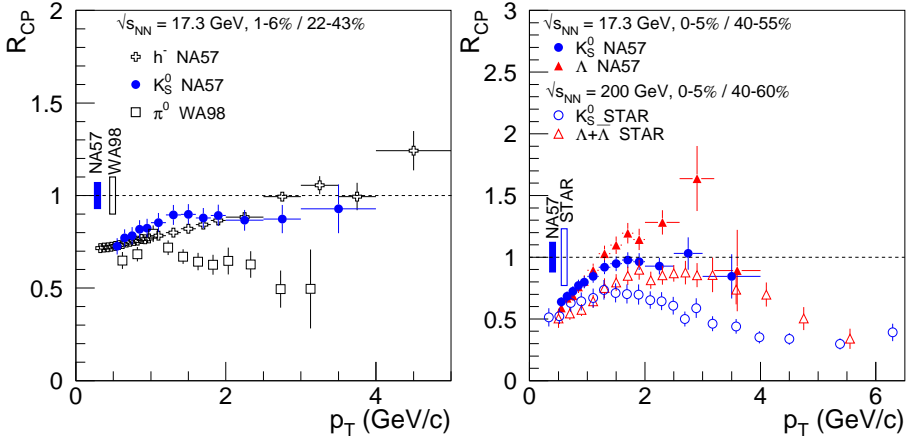}
   \caption{(colour online) Left: $\RCP(\pt)$ for $h^-$ and $\Kzs$ from NA57
            and $\pi^0$ from WA98~\cite{wa98pi0Rcp}
            in Pb--Pb collisions at $\sqrtsNN=17.3~\gev$.
            Right: $\RCP(\pt)$ for $\Kzs$ and $\Lambda$
            in Pb--Pb at $\sqrtsNN=17.3~\gev$ (NA57)
            and in Au--Au at $\sqrtsNN=200~\gev$
            (STAR)~\cite{starklaRcp}; slightly different
            peripheral classes are employed for the comparison.
            The bars
            centered at $\RCP=1$ represent the normalization
            errors;
            the point-by-point bars are the quadratic sum of statistical and
            systematic errors.}
   \label{fig:cmpexp}
\end{center}
\end{figure}

In Fig.~\ref{fig:cmpexp} we compare our results to
$\RCP$ measurements at the SPS and at RHIC. In the left-hand
panel, the WA98 $\pi^0$ data~\cite{wa98pi0Rcp} for the ratio
1--6\%/22--43\% in \PbPb~collisions at $\sqrtsNN=17.3~\gev$
are plotted together with the NA57 $h^-$ and $\Kzs$ data for the same
centrality classes\footnote{In the WA98 publication~\cite{wa98pi0Rcp},
the centrality classes are defined using percentiles of the measured
minimum bias cross section, 
$\sigma^{\rm Pb-Pb}_{\rm WA98\,m.b.}\approx 6.3~\rm b$.
Here, we rename the classes in terms of percentiles of the 
inelastic cross section, $\sigma^{\rm Pb-Pb}_{\rm inel}$.
The values used by the two experiments for the inelastic cross section
are very similar: 7.26~b for NA57 and 7.41~b for WA98.}. 
Using these classes, the $\Kzs$ $\RCP$ is approximately constant at 
0.9 for $\pt>1~\gev/c$ and is
significantly larger than that measured by the WA98 Collaboration 
for $\pi^0$ ($\RCP\approx 0.6$), 
even when taking into account the normalization
systematic errors, independent for the two experiments.
The $h^-$ data from NA57 are compatible, within the systematic errors, 
with the $\pi^0$ data from WA98 for $\pt\lsim 1.5~\gev/c$, where the $h^-$
sample is expected to be dominated by $\pi^-$. For higher $\pt$, 
$h^-$ have a larger $\RCP$ than $\pi^0$; this may be due to increasing 
contributions from $\rm K^-$ and $\rm\overline p$ in the $h^-$ sample.
At top RHIC energy, $\sqrtsNN=200~\gev$, the kaon $\RCP$ 
(as measured by PHENIX~\cite{phenixpikpRcp} and STAR~\cite{starklaRcp}) 
is larger than that of neutral pions 
(PHENIX~\cite{phenixpikpRcp}) for 
$\pt\lsim 2~\gev/c$, while they are similar for higher transverse momenta.
The observed difference in $\RCP$ between kaons and pions at SPS energy, 
and at RHIC energy for low $\pt$, is reminiscent of the `Cronin enhancement'
above $N_{\rm coll}$-scaling originally observed for $3<\pt<6~\gev/c$ 
in proton--nucleus (pA) collisions at $\sqrtsNN$ values up to 
$38.8~\gev$~\cite{cronin}. 
This enhancement, commonly interpreted 
as due to initial-state multiple scattering (partonic intrinsic transverse
momentum broadening), was in fact found to 
increase according to the hierarchy pions--kaons--protons~\cite{cronin}.
At RHIC energy, owing to the hardening of the $\pt$ distributions, 
the effect would be reduced compared to lower energies, as confirmed by 
preliminary results on the particle-species dependence of the \mbox{d--Au}
(similar to proton--nucleus) $\RCP$ 
at $\sqrtsNN=200~\gev$~\cite{phenixdAu}.

The comparison for $\Kzs$ and $\La$ at SPS and RHIC (STAR
data for \mbox{Au--Au} at $\sqrtsNN=200~\gev$~\cite{starklaRcp})
is presented in the right-hand panel of Fig.~\ref{fig:cmpexp}.
In the $\pt$ range covered by our data, up to $4~\gev/c$,
the {\em relative} pattern for $\Kzs$ and $\La$ is similar at
the two energies, while absolute values are higher at SPS than
at RHIC, where parton energy loss is believed to have a
strong effect. Part of the difference between the $\RCP$ values at the two 
energies may be due to different nuclear modification of the 
parton distribution functions (PDFs). 
In fact, in the $x_{\rm Bjorken}$ range relevant 
for parton production at a given $\pt$ 
at the SPS, e.g.\,$x\simeq 0.3$ for $\pt\simeq 3~\gev/c$, 
nuclear PDFs are expected to be enhanced by about 10--20\% 
(anti-shadowing), while almost no effect is expected for the smaller
values, $x\simeq 0.03$, relevant for the same $\pt$ values at RHIC energy 
(see e.g.\,\cite{eks}). 
At RHIC, the larger $\RCP$ for $\La$ with respect
to kaons~\cite{starklaRcp} or, more generally, for baryons with respect 
to mesons~\cite{phenixpikpRcp}, in
the intermediate $\pt$ range, $2$--$4~\gev/c$, has been interpreted
as due to parton coalescence~\cite{sorensen,recoth} in a
high-density medium with partonic degrees of freedom. Our data
show that a similar $\La$--$\rm K$ pattern is present also at
$\sqrtsNN=17.3~\gev$. We note that such a pattern may also be
explained in terms of larger Cronin effect for $\La$ with respect to 
kaons.

\begin{figure}[!t]
\begin{center}
   \includegraphics[width=\textwidth]{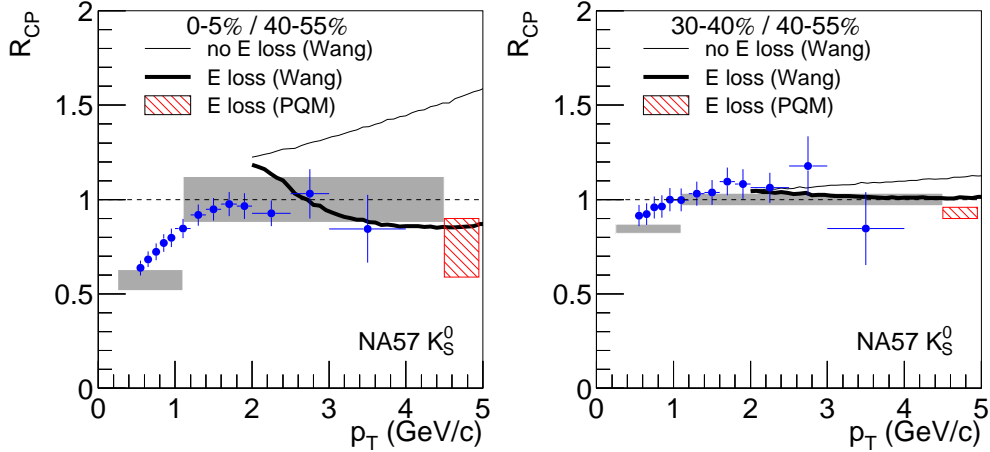}
   \caption{(colour online) $\RCP(\pt)$
            for $\Kzs$ in \PbPb~collisions at $\sqrtsNN=17.3~\gev$,
            compared to predictions~\cite{wangprivate,pqm}
            with and without
            the effect of parton energy loss (details in the text).}
   \label{fig:cmpth}
\end{center}
\end{figure}

In Fig.~\ref{fig:cmpth} we compare our $\Kzs$ data to predictions provided by
X.N.~Wang~\cite{wangprivate}, obtained from a perturbative-QCD-based
calculation~\cite{gyulassywang,wang}, including (thick line) or
excluding (thin line) in-medium parton energy loss. The initial
gluonic rapidity density of the medium, $\d N_{\rm g}/\d y$, 
was scaled down from that
needed to describe RHIC data, according to the decrease by
about a factor 2 in the charged multiplicity per unit of rapidity
from RHIC to SPS energy. For the
0--5\%/40--55\% $\RCP$, the curve without energy loss shows a
large enhancement, increasing with $\pt$. In the calculation this
enhancement arises principally from initial-state partonic intrinsic
transverse momentum broadening, which is assumed to be proportional to the
number of scatterings that the two colliding partons suffer inside the 
nuclei before the hard scattering, and thus larger for central 
than for peripheral collisions~\cite{wang}.
The magnitude of the broadening is tuned on the basis of the 
original Cronin effect data~\cite{cronin}.
A `Cronin-like' enhancement is observed in the \mbox{d--Au}
central-to-peripheral nuclear modification factor 
for charged kaons at RHIC energy~\cite{phenixdAu}, 
while it is not present in our
nucleus--nucleus $\Kzs$ data. The curve that includes  energy
loss, scaled down from RHIC as explained above, describes the data
better. Moving to less central collisions the predictions with and
without energy loss get closer to one another, and both are
compatible with the data. As a cross-check, we compare the $\RCP$
value predicted by X.N.~Wang with energy loss to the prediction of an
independent model of parton energy loss, the Parton Quenching
Model (PQM), based on the BDMPS formalism~\cite{bdmps},
that describes several energy-loss-related observables at RHIC
energies~\cite{pqm}. Since the PQM model does not include initial-state
effects, the predicted $\RCP$ was rescaled using the Wang baseline
without energy loss (thin line). The PQM result has an 
uncertainty due to the fact that the medium-induced energy loss
becomes of the order of the initial parton energy~\cite{pqm}; 
since this uncertainty is larger for low-energy partons, the 
calculation result is meaningful for $\pt\gsim 4~\gev/c$ only.
The PQM result, shown in Fig.~\ref{fig:cmpth} by the hatched areas, agrees with
the Wang calculation with energy loss (thick line).
The two models predict a similar energy loss
effect at SPS energy, i.e.\,a reduction  of
the 0--5\%/40--55\% $\RCP$ (for $\pt\gsim 4~\gev/c$)
by about a factor 2, with respect to the value calculated without 
energy loss.

\section{Conclusions}
\label{conclusions}

Central-to-peripheral nuclear modification factors
for $\Kzs$, $\La$, $\Al$ and $h^-$ in \mbox{Pb--Pb} collisions at
top SPS energy have been measured as a function of $\pt$ up to 
about $4~\gev/c$.
At low $\pt$, $\RCP$ agrees with $N_{\rm part}$ scaling for all the
particles under consideration, except the $\Al$, for which the yields 
at low $\pt$ are found to increase slower than the number of
participants.
For $\pt\gsim 1~\gev/c$, $\Kzs$, $\La$ and $\Al$ show a 
pattern similar to that observed in \mbox{Au--Au} collisions 
at top RHIC energy, although the $\RCP$ values are found to be larger at SPS.
At RHIC, this pattern has been interpreted in the framework of 
models that combine 
parton energy loss with hadronization via coalescence, at intermediate $\pt$, 
and via fragmentation, at higher $\pt$.
The measured $\Kzs$ 0--5\%/40--55\% $\RCP$
is not reproduced by a theoretical calculation that includes only
initial-state nuclear effects.
The data can be better described by including
 final-state parton energy loss as predicted for SPS energy
on the basis of RHIC data.

\paragraph*{Acknowledgments.}
We thank V.~Greco, X.N.~Wang and U.A.~Wiedemann for fruitful discussions
and T.~Peitzmann for making available to us the WA98 data.

\end{document}